\documentclass[prl,twocolumn, amsmath, amssymb, showpacs]{revtex4-1}
\newcommand{\formula}{${\rm LiCu_{2}O_{2}}\;$}
\usepackage{graphicx}
\usepackage{dcolumn}
\usepackage{bm}
\usepackage[dvipdfm,
            colorlinks,linkcolor=blue,
            citecolor=blue,
            hyperindex,
            pdfstartview=FitH,
            plainpages=false]
            {hyperref}

\begin{document}

\preprint{ }

\title{Temperature-Dependent Anomalies in the Structure of the (001) Surface of \formula}

\author{Xuetao Zhu$^1$, Yangyang Yao$^1$, H. C. Hsu$^2$, F. C. Chou$^2$ and M. El-Batanouny$^1$}

\affiliation{\sl{$^1$ Physics Department, Boston University, Boston,
MA 02215\\$^2$ Center of Condensed Matter Sciences, National Taiwan
University, Taipei 10617, Taiwan}}
\date{\today}

\begin{abstract}
Surface corrugation functions, derived from elastic helium atom
scattering (HAS) diffraction patterns at different temperatures,
reveal that the Cu$^{2+}$ rows in the (001) surface of \formula
undergo an outward displacement of about 0.15 {\AA} as the surface
was cooled down to 140 K. This is probably the first time that
isolated one-dimensional magnetic ion arrays were realized, which
qualifies the $\rm Li^{1+}Cu^{2+}O^{2-}_2$ surface as a candidate to
study one-dimensional magnetism. The rising Cu$^{2+}$ rows induce a
surface incommensurate structural transition along the
$a$-direction. Surface equilibrium analysis showed that the surface
Cu$^{2+}$ ions at bulk-like positions experience a net outward force
along the surface normal which is relieved by the displacement.
Temperature-dependent changes of the surface phonon dispersions
obtained with the aid of inelastic HAS measurements combined with
surface lattice dynamical calculations are also reported.
\end{abstract}


\maketitle

Over the past decade \formula has attracted considerable attention
because it incorporates double-chain ladders of Cu$^{2+}$O, which
makes it a prototypical quasi-one-dimensional quantum spin-1/2
magnetic system with competing magnetic interactions \cite{Zvyagin,
Choi, Masuda1, Gippius}. It still remains an exciting system since
the recent discovery of ferroelectricity induced by an ordered
helimagnetic phase,  making  \formula the second cuprate to join the
list of multiferroics \cite{Park, Xiang, Seki, Huang}; $\rm
LiCuVO_4$ \cite{Naito} being the first.

It was reported that \formula exhibits a spin-singlet (or a gapped
spin-liquid) ground state for temperatures $T>30$ K \cite{Zvyagin}.
Two successive magnetic phase transitions bring the system into an
incommensurate long-range ordered state with a helicoidal spin
structure in which the arms of the double-chain are
antiferromagnetically ordered
\cite{Masuda1,Huang,Seki,Rusydi,Yasui,Kobayashi}. The picture that
emerges reveals the presence of a collinear sinusoidal spin-ordered
phase, with spin polarization along the $c$-axis, below the
transition temperature $T_N\simeq 24.6$ K. An initial modulation
wave vector ${\bf Q}=(0,0.172,0)$ r.l.u. was reported, and was found
to increase with decreasing temperature. At the second transition
temperature $T_{FE}\simeq 23.0$ K, the spin polarization acquires
small components along the $a$ and $b$ axes, giving rise to the
helicoidal spin-ordered structure. This structure is characterized
by an ellipse with the helical axis tilted by $45^{\circ}$ from the
$b$-axis within the $ab$ plane, and with the incommensurate
modulation still along the $b$-axis \cite{Rusydi, Kobayashi}. ${\bf
Q}$ continues to increase and seems to saturate at $T\simeq 12$ K
with ${\bf Q}=(0,0.174,0)$. The onset of the helicoidal phase
induces ferroelectricity with polarization along the $c$-axis.
Moreover, recent studies by resonant soft x-ray magnetic scattering
\cite{Huang} have demonstrated that the long-range ordered magnetic
states are actually 2D-like ground states, where the 2D character
was attributed to a small but effective spin coupling along the
$c$-axis, which, in turn, suppresses quantum fluctuations. These
measurements also revealed the presence of short-range spin
correlations above $T_N$, which eventually disappear at about $30$ K
leaving a spin liquid ground state at higher temperatures
\cite{Huang}.

The onset of ferroelectricity below $T_{FE}$ in \formula was first
reported by Park {\it et al.} \cite{Park}. Several models were
proposed to explain the origin of electrical polarization in
helicoidal magnetic systems. Two microscopic models were introduced
to explain the mechanism responsible for multiferroicity in such
systems, the first is based on the relation between the
magnetoelectric effect and microscopic spin currents \cite{Katsura}
and the second invoked the idea of inverse Dzyaloshinskii-Moriya
\cite{Sergienko}. In the former, the polarization is given by
$$\mathbf{P}\propto\mathbf{e}_{ij}\times(\mathbf{S}_i
\times\mathbf{S}_j),$$ where $\mathbf{S}_i$ and $\mathbf{S}_j$ are
the local spins at site $i$ and site $j$ respectively, and
$\mathbf{e}_{ij}$ is the unit vector connecting the $i$ and $j$
sites \cite{Katsura}. Subsequently, a phenomenological model based
on electric polarization related symmetry-invariant Landau-Ginzburg
terms of the form $$\frac{{\bf P}^2}{2\chi_2}+\gamma{\bf
P}\boldsymbol{\cdot}\left[{\bf
M}\,\left(\boldsymbol{\nabla\cdot}{\bf M}\right)-\left({\bf
M}\boldsymbol{\cdot\nabla}\right)\,{\bf M}+\ldots\right]$$ was
proposed \cite{Mostovoy}. In the case of \formula all these models
consistently support a spiral spin with components in the
$bc$-plane, which agrees with the findings of Ref. \cite{Seki,
Rusydi,Huang,Yasui,Kobayashi}.

Despite the extensive studies of bulk structural and magnetic
properties of \formula crystals cited above, to our knowledge, the
only investigation of its surface properties was reported by the
current authors \cite{Yao}, and involved helium atom scattering
(HAS) studies of the structure and dynamics of the (001) surface of
\formula at room temperature. In that work, it was shown that the
surface termination is exclusively a $\rm Li^{1+}Cu^{2+}O^{2-}_2$
plane. In this letter we report a surprising discovery of
temperature-dependent anomalies observed on this surface. For
temperatures below 200 K, the $\rm Cu^{2+}$ rows were found to rise
above the surface plane; this, in turn, induces an incommensurate
surface structure. As far as we know, this is probably the first
time that an isolated one-dimensional row of magnetic ions was
realized.

\formula has a layered charge-ordered orthorhombic crystal structure
belonging to the {\sl Pnma} space group. The primitive cell has
lattice constants $a$=5.73, $b$=2.86, and $c$=12.47 $\rm\AA$. Single
crystals of ${\rm Li_xCu_{2}O_{2}}$ with Li content of
$x\sim0.99\pm0.03$, which can be considered as stoichiometric
\formula, were grown by the floating-zone process \cite{Hsu}. The
experimental setup and procedures of the HAS measurements are
described in details in Ref. \cite{Yao, Farzaneh}.

In this work, we present surface diffraction patterns of the $\rm
Li^{1+}Cu^{2+}O^{2-}_2$ termination, obtained by the elastic HAS at
temperatures below 210 K. Analysis of diffraction intensities
yielded surface corrugations, which can be considered as direct
pictures of the geometrical arrangement of the surface atoms
\cite{Rieder}. As we previously reported \cite{Yao}, diffraction
patterns recorded between room temperature and roughly 230 K showed
none of the anomalies discussed below.

\begin{figure}[h!]
\includegraphics[width=0.5\textwidth]{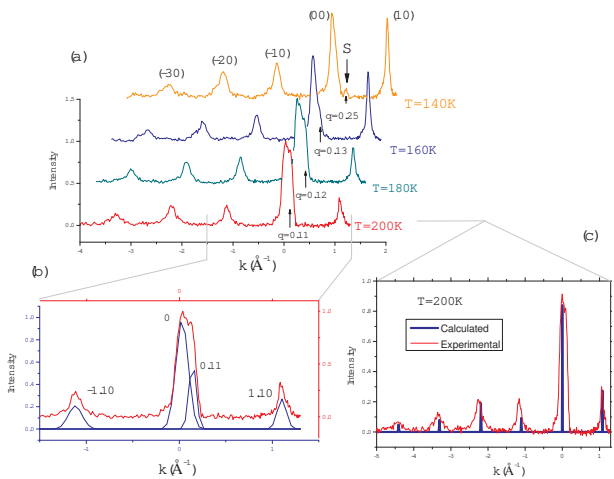}\\\vspace{-0.1in}
\caption{(a) Diffraction patterns along the $a$-direction obtained
at $T=200,\,180,\,160$ and $140$ K. Intensities are normalized to a
specular intensity of 1. Arrows indicate the presence of satellite
peak, denoted by $S$, at the specified wave vectors q. (b) Gaussian
fit (blue) to experimental peaks obtained at 200 K. (c) Comparison
between the measured diffraction peaks (red) and the calculated ones
(blue). See text for details about the calculation.}\label{fig1}
\end{figure}

\begin{figure}[h!]
\includegraphics[width=0.23\textwidth]{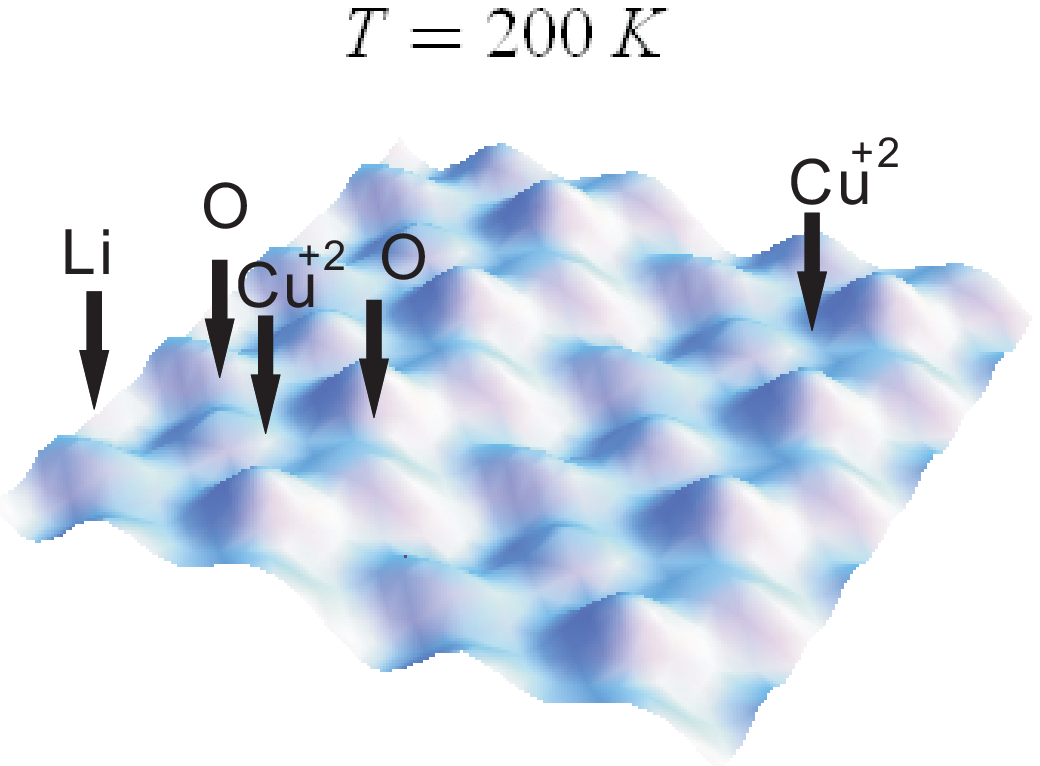}
\includegraphics[width=0.23\textwidth]{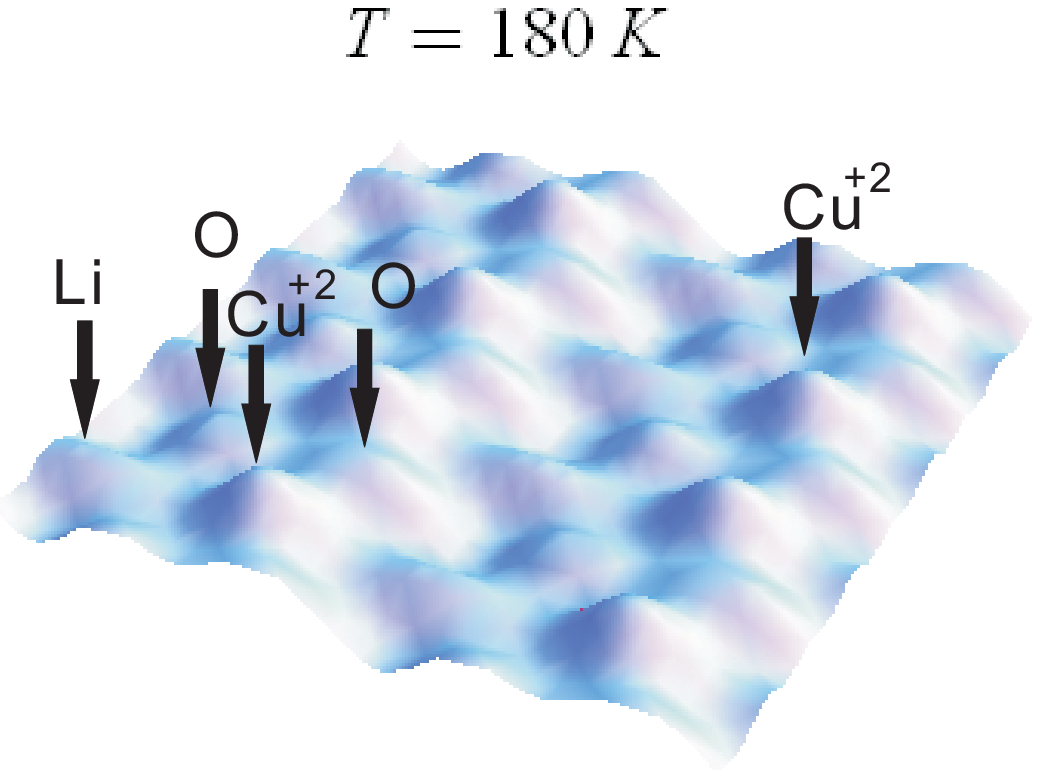}
\includegraphics[width=0.23\textwidth]{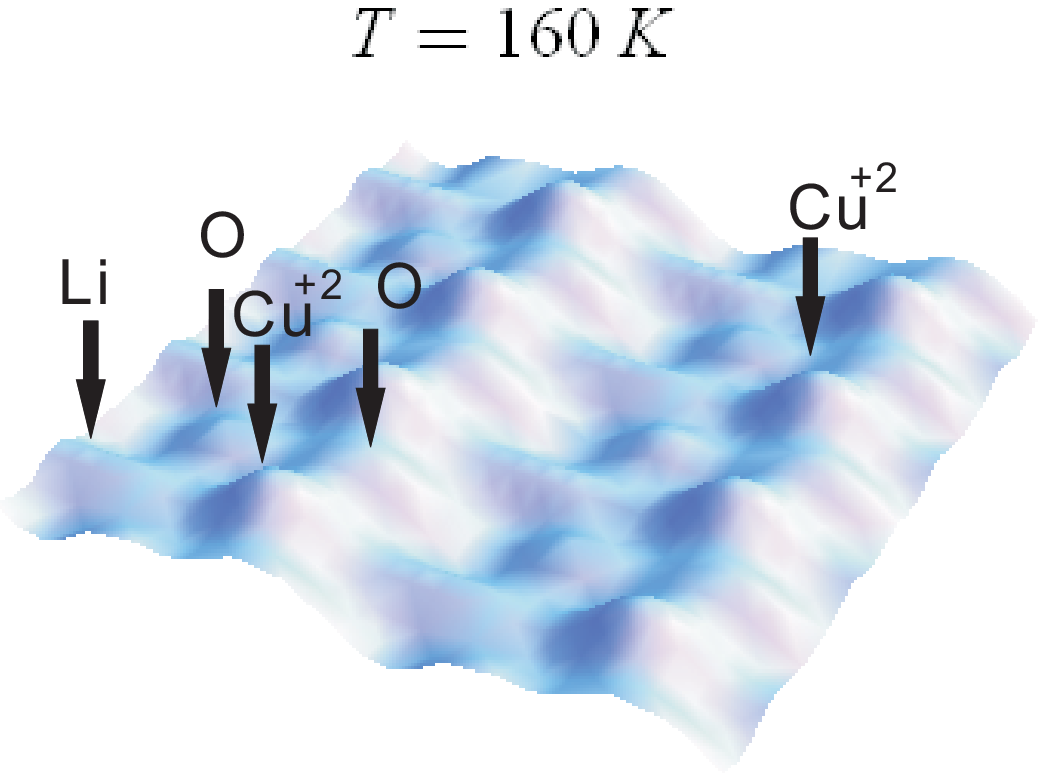}
\includegraphics[width=0.23\textwidth]{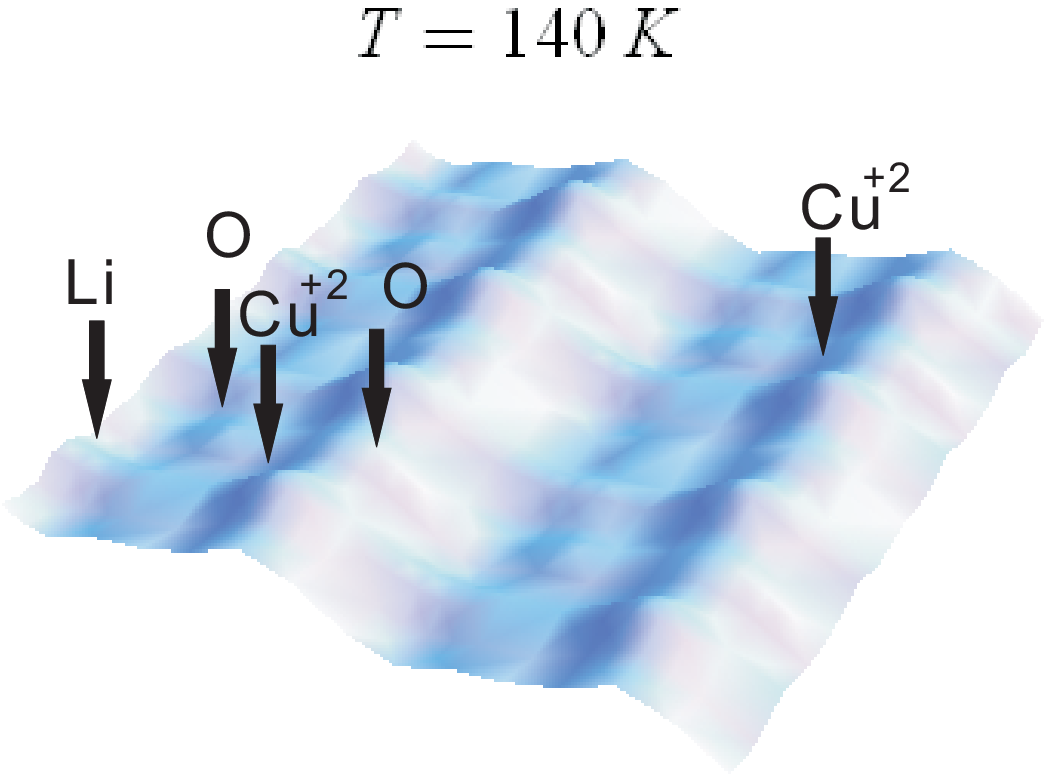} \\\vspace{-0.2in}\caption{Surface corrugations at
different temperatures with $T$ decreasing.}\label{fig2}
\end{figure}

\formula samples were cleaved in situ, under ultra-high-vacuum
conditions at $T=200$ K. Subsequently, the samples were cooled down
in steps to $T=140$ K. Fig. \ref{fig1}(a) shows four diffraction
patterns along the $a$-direction ($\langle 10 \rangle$-direction)
recorded at 200, 180, 160 and 140 K, respectively. It is clear that
the intensity of the (10) first-order peak increases with decreasing
temperature. Moreover, a shoulder is observed on the right side of
the (00) specular peak at 200 K and progressively separates from it
with decreasing temperature. A Gaussian fit to the 200 K diffraction
peaks (DPs), Fig. \ref{fig1}(b), shows that the shoulder is actually
a satellite peak at ${\bf q}=(0.11,0)\ {\rm \AA}^{-1}$. Similar fits
to the DPs at lower temperatures show that ${\bf q}$ increases
monotonically with decreasing temperature to ${\bf q}=(0.25,0)\ {\rm
\AA}^{-1}$ at 140 K. Henceforth, we denote this satellite peak by
$S$. Further cooling to temperatures below 140 K showed no change in
the diffraction pattern.

We used the \textit{hard corrugated wall model} within the
\textit{eikonal approximation} to calculate the elastic scattering
intensities and followed an iterative computational method to fit
the calculated intensities to experimental measurements. Details of
this iterative computational method can be found in Ref. \cite{Yao}
and references therein. Fig. \ref{fig1}(c) shows a very good match
between the experimental diffraction pattern at 200 K (red curve)
and the calculated intensities of the DPs (blue rods). An important
product of this best fit is the topology of the primitive cell
manifest in the two-dimensional surface corrugation function (SCF).
The SCFs at $T=200$ , $180$, $160$ and $140$ K are shown in Fig.
\ref{fig2}, where the surface ion positions are indicated by arrows.
It is clear from the SCFs that the Cu$^{2+}$ ions rise monotonically
with decreasing temperature. By calculating the difference between
the maxima and minima of the SCFs, the largest displacement of the
Cu$^{2+}$ ions is estimated to be 0.14-0.15 {\AA} above their
bulk-like surface positions.

The presence of the satellite peak, S, is a manifestation of a
surface incommensurate structure (SIS) along the $a$-direction.
There can be two possible causes for the onset of the SIS: First,
the appearance of surface electric dipoles associated with the
elevation of the Cu$^{2+}$ rows renders the surface unstable.
Second, the tendency to maintain Cu$^{2+}-$O$^{2-}$ bond length
close to its original value induces a lateral displacement of
neighboring rows of O$^{2-}$ ions toward the Cu$^{2+}$ rows along
the $a$-direction.

\begin{figure}[h!]
\includegraphics[width=0.5\textwidth]{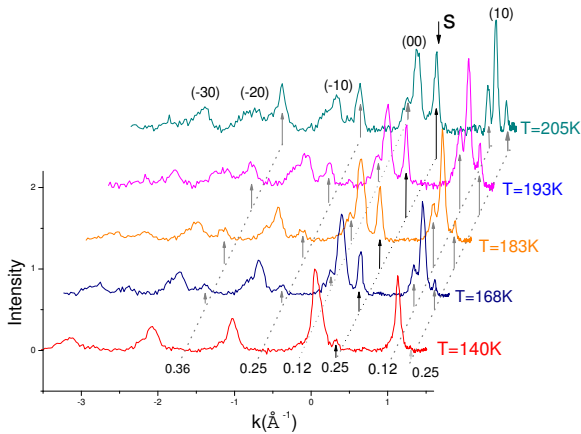}\\\vspace{-0.2in}
\caption{Evolution of diffraction patterns along the $a$-direction
with temperature increase above 140 K. Intensities are normalized to
the specular intensity. Arrows mark satellite peak positions with
their $q$ values.}\label{fig3}
\end{figure}

Fig. \ref{fig3} shows five diffraction patterns along the
$a$-direction sampled as the temperature was raised after cooling to
140 K. A sequence of satellite peaks emerges, signaling the
development of the SIS phases. The locations of these satellites are
marked by arrows. The corresponding wave vectors, determined by
Gaussian fits, are recorded next to each arrow. It should be noticed
that the satellite intensities increase with increase in
temperature, but their wave vectors remain constant. The emerging
SIS is characterized by satellite wave vectors: ${\bf
q}=(0.12\pm0.01,\ 0)$, ${\bf q}=(0.25\pm0.01,\ 0)$ and ${\bf
q}=(0.36\pm0.02,\ 0)\ {\rm \AA}^{-1}$. An incommensurability close
to $9/8$ was obtained by comparison with the commensurate first
order wave vector of ${\bf q}=(1.10,\ 0)\ {\rm \AA}^{-1}$, in Fig.
\ref{fig1}(b). All features associated with the incommensurate phase
and the rise of the Cu$^{2+}$ rows in the diffraction pattern
completely disappear above ~220 K.

\begin{figure}[h!]
\includegraphics[width=0.45\textwidth]{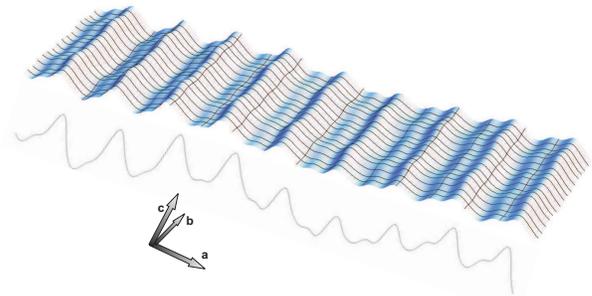}\\\vspace{-0.2in}
\caption{Surface corrugation with a cross cut line for the
incommensurate structure.}\label{fig4}
\end{figure}

\begin{figure}[h!]
\includegraphics[width=0.48\textwidth]{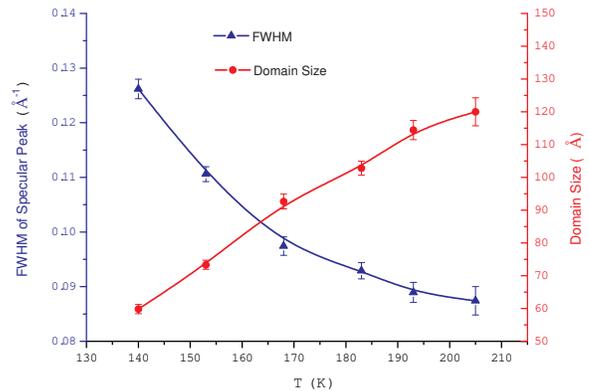}\\\vspace{-0.2in}
\caption{Full-width-half-maximum ($\Delta k$) of the specular peak
(blue), as well as the deconvoluted average domain size (red) as a
function of temperature increase.}\label{fig5}
\end{figure}

Employing the approximate periodicity of $9/8$, we carried out a
peak intensity fitting procedure similar to that described above.
The resulting SCF is shown in Fig. \ref{fig4}. Moreover, we
extracted the full-width-half-maximum (FWHM) of the specular peak
from the diffraction patterns of Fig. \ref{fig3}; they are plotted
in Fig. \ref{fig5}, together with the corresponding average domain
size, as a function of temperature. The latter was obtained by
deconvoluting the specular peak with an instrument transfer width
\cite{Park2} of $\Delta k_{inst}=0.07\ $\AA$^{-1}$. It is clear from
Fig. \ref{fig3} and \ref{fig5} that the incommensurate domains grow
with rise in temperature.

\begin{figure}[h!]
\includegraphics[width=0.43\textwidth]{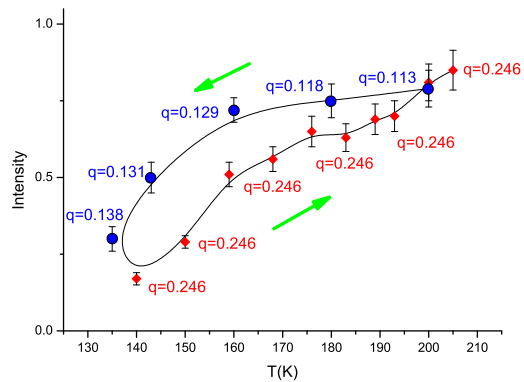}\\\vspace{-0.2in}
\caption{The intensity of the $S$ satellite peak. All intensities
are normalized by setting the specular intensity to 1. The blue dots
indicate the evolution of the corresponding $q$-vector with
decreasing temperature. The red dots mark the evolution of the
$S$-peak intensity at the saturated q-vector, as obtained upon
heating. The black line is a guide to eye and green arrows show the
temperature increasing or decreasing procedure.}\label{fig6}
\end{figure}

In Fig. \ref{fig6} we plot the intensity of the $S$-satellite peak
at different temperatures, recorded during the cooling and heating
segments of the experimental measurements. It is clear from Fig.
\ref{fig6} that the temperature dependence of the $S$-peak intensity
displays a hysteresis loop with respect to the two segments of
cooling and heating. The overall behavior points to a first-order
commensurate-incommensurate phase transition.

In order to understand the experimental observations outlined above,
we investigated the origin of the anomalous outward displacement of
the Cu$^{2+}$ rows with the aid of detailed surface equilibrium
analysis \cite{Boyer}. This study showed that at bulk-like positions
the surface Cu$^{2+}$ ions experience a net outward force; and that
this force can be relieved by displacing the Cu$^{2+}$ rows outward
from their bulk positions, along the surface normal \cite{Yao}.

\begin{figure}[h!]
\includegraphics[width=0.5\textwidth]{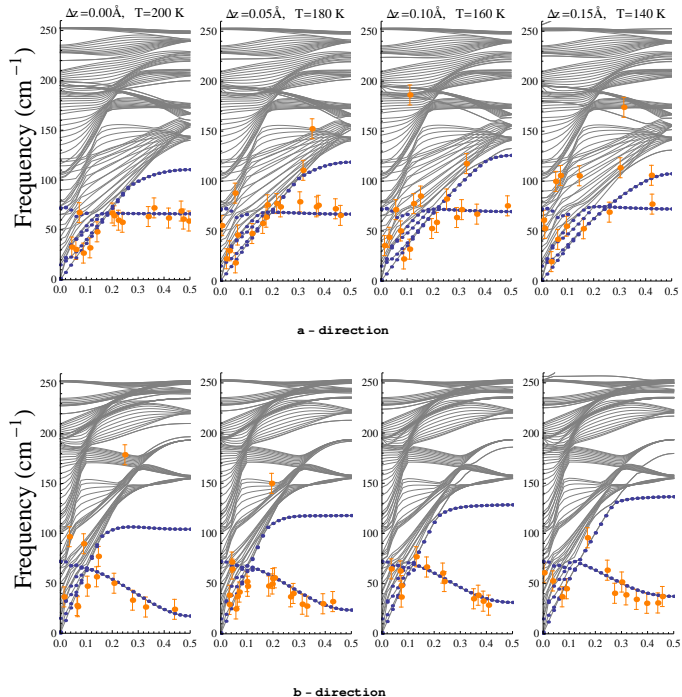}\\\vspace{-0.2in}
\caption{Measured inelastic events (orange dots with error bars)
reduced to the proper surface Brillouin zone (SBZ), as a function of
decreasing temperature, superposed on calculated phonon dispersion
curves corresponding to different surface Cu$^{2+}$ displacements
($\Delta z$). The calculated surface phonon modes are indicated by
blue dots, and the gray background represents the projection of the
bulk bands on the SBZ. Upper panels: a-direction. Lower panels:
b-direction.}\label{fig7}
\end{figure}

Finally, we studied the changes in the surface phonon dispersion as
a function of decreasing temperature. In these studies we employed
inelastic HAS, together with lattice dynamical calculations based on
a slab model with variable surface Cu$^{2+}$ positions. The
procedural details of these studies can be found in Ref. \cite{Yao,
Farzaneh}. Fig. \ref{fig7} shows the results along high-symmetry
directions - the experimental inelastic events are indicated by
orange dots with error bars. The measurements were carried out at
$T=200$, $180$, $160$ and $140$ K, and the surface phonon dispersion
curves were calculated at different surface Cu$^{2+}$ displacements
from bulk-like positions, indicated by $\Delta z$, which were
obtained from corresponding SCFs. As was reported in Ref.
\cite{Yao}, the lowest two surface phonon dispersion branches (blue
dots in Fig. \ref{fig7}) involve the motion of Cu$^{2+}$ and
Li$^{1+}$ ions normal to the surface. We notice the following trends
with decrease in temperature: the lower branch along the
$b$-direction displays a gradual increase in frequency close to the
surface Brillouin zone (SBZ) boundary, while the slope of the upper
branch and its SBZ frequency gradually decrease along the
$a$-direction.

In summary, we presented experimental evidence that the Cu$^{2+}$
rows in the (001) surface of \formula undergo an outward
displacement of about 0.15 {\AA} as the surface is cooled down to
140 K; and that this displacement induces an incommensurate
structure along the $a$-direction. The growth of the incommensurate
domains was found to be thermally activated. The rise of the
Cu$^{2+}$ rows was supported by surface equilibrium analysis, which
showed that the bulk-like positions experienced outward forces that
were relieved by the displacement. Low-lying surface phonon
branches, associated with the motion of Cu$^{2+}$ and Li$^{1+}$ ions
normal to the surface, exhibit stiffening along the $b$-direction
and softening along the $a$-direction as the temperature is
lowered.\\

M. El-Batanouny acknowledges support from the U.S. Department of
Energy under Grant No. DE-FG02-85ER45222. F. C. Chou acknowledges
support from National Science Council of Taiwan under project No.
NSC-95-2112-M-002. ME would like to thank C. Chamon and W. Klein for
valuable discussions.

\end{document}